\documentclass[aps,prl,twocolumn,showpacs,superscriptaddress,floatfix]{revtex4}
\def\FigFactor{0.475}

\usepackage{graphics}

\def\pT{\mbox{$p_T$}}

\def\sqrtsNN{\mbox{$\sqrt{s_\mathrm{NN}}$}}
\def\sqrts{\mbox{$\sqrt{s}$}}

\def\Nbinary{\mbox{$\mathrm{N}_\mathrm{bin}$}}
\def\NbinaryMean{\mbox{$\langle\Nbinary\rangle$}}

\def\pbar{\mbox{$\bar\mathrm{p}$}}

\def\hplus{\mbox{$\mathrm{h}^+$}}
\def\hminus{\mbox{$\mathrm{h}^-$}}
\def\hphm{\mbox{$(\hplus+\hminus)/2$}}

\def\RAB{\mbox{$R_{AB}(\pT)$}}

\def\TAB{\mbox{$T_{AB}$}}

\def\sigmappinel{\mbox{$\sigma^{pp}_{inel}$}}

\def\sigmadAuhadr{\mbox{$\sigma^{dAu}_{hadr}$}}

\def\lt{\mbox{$<$}}
\def\gt{\mbox{$>$}}

\begin{document}

\title{Evidence from d+Au measurements for final-state suppression 
of high \pT\ hadrons in Au+Au collisions at RHIC}


\affiliation{Argonne National Laboratory, Argonne, Illinois 60439}
\affiliation{Brookhaven National Laboratory, Upton, New York 11973}
\affiliation{University of Birmingham, Birmingham, United Kingdom}
\affiliation{University of California, Berkeley, California 94720}
\affiliation{University of California, Davis, California 95616}
\affiliation{University of California, Los Angeles, California 90095}
\affiliation{Carnegie Mellon University, Pittsburgh, Pennsylvania 15213}
\affiliation{Creighton University, Omaha, Nebraska 68178}
\affiliation{Nuclear Physics Institute AS CR, \v{R}e\v{z}/Prague, Czech Republic}
\affiliation{Laboratory for High Energy (JINR), Dubna, Russia}
\affiliation{Particle Physics Laboratory (JINR), Dubna, Russia}
\affiliation{University of Frankfurt, Frankfurt, Germany}
\affiliation{Indiana University, Bloomington, Indiana 47408}
\affiliation{Insitute  of Physics, Bhubaneswar 751005, India}
\affiliation{Institut de Recherches Subatomiques, Strasbourg, France}
\affiliation{University of Jammu, Jammu 180001, India}
\affiliation{Kent State University, Kent, Ohio 44242}
\affiliation{Lawrence Berkeley National Laboratory, Berkeley, California 94720}
\affiliation{Max-Planck-Institut f\"ur Physik, Munich, Germany}
\affiliation{Michigan State University, East Lansing, Michigan 48824}
\affiliation{Moscow Engineering Physics Institute, Moscow Russia}
\affiliation{City College of New York, New York City, New York 10031}
\affiliation{NIKHEF, Amsterdam, The Netherlands}
\affiliation{Ohio State University, Columbus, Ohio 43210}
\affiliation{Panjab University, Chandigarh 160014, India}
\affiliation{Pennsylvania State University, University Park, Pennsylvania 16802}
\affiliation{Institute of High Energy Physics, Protvino, Russia}
\affiliation{Purdue University, West Lafayette, Indiana 47907}
\affiliation{University of Rajasthan, Jaipur 302004, India}
\affiliation{Rice University, Houston, Texas 77251}
\affiliation{Universidade de Sao Paulo, Sao Paulo, Brazil}
\affiliation{University of Science \& Technology of China, Anhui 230027, China}
\affiliation{Shanghai Institute of Nuclear Research, Shanghai 201800, P.R. China}
\affiliation{SUBATECH, Nantes, France}
\affiliation{Texas A \& M, College Station, Texas 77843}
\affiliation{University of Texas, Austin, Texas 78712}
\affiliation{Valparaiso University, Valparaiso, Indiana 46383}
\affiliation{Variable Energy Cyclotron Centre, Kolkata 700064, India}
\affiliation{Warsaw University of Technology, Warsaw, Poland}
\affiliation{University of Washington, Seattle, Washington 98195}
\affiliation{Wayne State University, Detroit, Michigan 48201}
\affiliation{Institute of Particle Physics, CCNU (HZNU), Wuhan, 430079 China}
\affiliation{Yale University, New Haven, Connecticut 06520}
\affiliation{University of Zagreb, Zagreb, HR-10002, Croatia}

\author{J.~Adams}\affiliation{University of Birmingham, Birmingham, United Kingdom}
\author{C.~Adler}\affiliation{University of Frankfurt, Frankfurt, Germany}
\author{M.M.~Aggarwal}\affiliation{Panjab University, Chandigarh 160014, India}
\author{Z.~Ahammed}\affiliation{Purdue University, West Lafayette, Indiana 47907}
\author{J.~Amonett}\affiliation{Kent State University, Kent, Ohio 44242}
\author{B.D.~Anderson}\affiliation{Kent State University, Kent, Ohio 44242}
\author{M.~Anderson}\affiliation{University of California, Davis, California 95616}
\author{D.~Arkhipkin}\affiliation{Particle Physics Laboratory (JINR), Dubna, Russia}
\author{G.S.~Averichev}\affiliation{Laboratory for High Energy (JINR), Dubna, Russia}
\author{S.K.~Badyal}\affiliation{University of Jammu, Jammu 180001, India}
% ...

\collaboration{STAR Collaboration}\homepage{www.star.bnl.gov}\noaffiliation


\date{\today; }

\begin{abstract}
We report measurements of single-particle inclusive spectra and
two-particle azimuthal distributions of charged hadrons at high
transverse momentum (high \pT) in minimum bias and central d+Au
collisions at \sqrtsNN=200 GeV. The inclusive yield is enhanced in d+Au collisions
relative to binary-scaled p+p collisions, while the two-particle
azimuthal distributions are very similar to those observed in p+p
collisions. These results demonstrate that the strong
suppression of the inclusive yield and back-to-back correlations at
high \pT\ previously observed in central Au+Au collisions are due to
final-state interactions with the dense medium generated in such
collisions.
\end{abstract}


\maketitle


Energetic partons propagating through matter
are predicted to lose energy through
induced gluon radiation, with the magnitude of the energy loss
depending strongly on the color charge density
\cite{EnergyLoss}. Partonic energy loss is potentially a 
sensitive probe of the matter created in high energy heavy-ion
collisions, where a quark-gluon plasma may form if sufficiently high
energy density is achieved. The energetic partons originate in the
hard scattering of partons from the incoming nuclei. Direct
measurement of jets resulting from parton fragmentation is difficult
in nuclear collisions; nevertheless partonic energy loss can be
studied using observables such as inclusive spectra and two-particle
azimuthal distributions of high transverse momentum (high \pT)
hadrons.

Measurements of high \pT\ hadron production in ultrarelativistic
interactions of heavy nuclei reveal strong suppression of both the
single-particle inclusive yield
\cite{STARHighpt130,PHENIXHighpt130,PHOBOSHighpt200,STARHighpt200} and
back-to-back pairs (large azimuthal separation $\Delta\phi$) in the
most-central, violent collisions, while near-side pairs (small
$\Delta\phi$) exhibit jet-like correlations that are similar to those
in proton+proton (p+p) collisions \cite{STARHighptCorrelations}. One
interpretation of these results is that, in the final state following
the hard scattering, energetic partons traversing the dense medium in the core of
the collision lose energy, and the observed jets are primarily those
created from partons produced near the surface and directed outwards
\cite{STARHighptCorrelations}. Alternatively, the suppression might
result from initial-state effects prior to the hard scattering, such
as the saturation of gluon densities in the incoming nuclei
\cite{KLM}. Models incorporating either picture 
are capable of describing central Au+Au collision data
\cite{STARHighpt200}.
Initial- and final-state effects in Au+Au collisions can be separated
through studies of deuteron(d)+Au collisions. Theoretical expectations
for d+Au collisions at the Relativistic Heavy Ion Collider (RHIC) are given
in~\cite{KLM,WangdAu,LevaidAu,KopeliovichdAu,Accardi,VitevdAu,Eskola,SpencerShadow,BaierEtalSaturation}.
Within a perturbative QCD (pQCD) framework, the expected initial-state
nuclear effects in d+Au collisions are multiple scattering prior to a
hard collision, which has been used to explain the Cronin enhancement of the
inclusive yield \cite{Antreysan}, and shadowing of the parton distribution
functions. Nuclear effects are expected to increase for
more central collisions; thus the centrality dependence of observables
measured in d+Au collisions also will help reveal their origin.

The STAR Collaboration reports
measurements of the inclusive
invariant \pT\ distribution and two-particle azimuthal distributions
at high \pT\ for charged hadrons [\hphm, approximated by the summed
yields of primary $\pi^\pm$, K$^\pm$, p and \pbar] in minimum bias and
central d+Au collisions at center of mass energy
\sqrtsNN=200 GeV per nucleon pair. Comparison is made to 
measurements at \sqrtsNN=200 GeV in the same detector for Au+Au and p+p interactions
\cite{STARHighpt200,STARHighptCorrelations}. The
inclusive yield is enhanced in d+Au collisions relative to
binary-scaled p+p
collisions, in contrast to the large suppression observed in central
Au+Au interactions.
Similar results are reported in \cite{PHENIXdAu,PHOBOSdAu, BRAHMSdAu}.
The d+Au two-particle azimuthal distributions
are very similar to those observed in p+p collisions. These
observations are consistent with expectations from pQCD models
incorporating both the Cronin enhancement and nuclear
shadowing~\cite{WangdAu,LevaidAu,KopeliovichdAu,Accardi,VitevdAu}, and
are inconsistent with calculations that attribute the suppression in
central Au+Au collisions to initial-state gluon
saturation~\cite{KLM}.


STAR is a multi-purpose detector \cite{STARNIM} located at Brookhaven National Laboratory's RHIC facility. 
For these measurements, the minimum bias trigger required at least one
beam-rapidity neutron in ZDC-Au, the Zero Degree Calorimeter (ZDC) in
the Au beam direction, which is assigned negative pseudorapidity ($\eta$).  This
trigger accepts 95$\pm$3\% of the d+Au hadronic cross section
\sigmadAuhadr. Trigger backgrounds were measured using beam bunches not in collision. 
Charged particle momenta were measured by the Time Projection Chamber
in a 0.5\,T solenoidal magnetic field.

After event selection cuts, the data set consists of $10^7$ minimum
bias d+Au events. Data were
analyzed using the techniques described in
\cite{STARHighpt130,STARHighptCorrelations}.  The vertex was reconstructed
in 93$\pm$1\% of triggered
minimum bias events. The spectra were
corrected for trigger and vertex-finding efficiencies. Contamination
of the spectra due to weak decay products was corrected based on
HIJING~\cite{HIJING}. Results of an independent analysis, using a
different technique for vertex reconstruction
\cite{STARHighpt200}, agree with the reported spectrum within the relative 
normalization uncertainties at all \pT.


\begin{figure}
\resizebox{\FigFactor\textwidth}{!}{\includegraphics{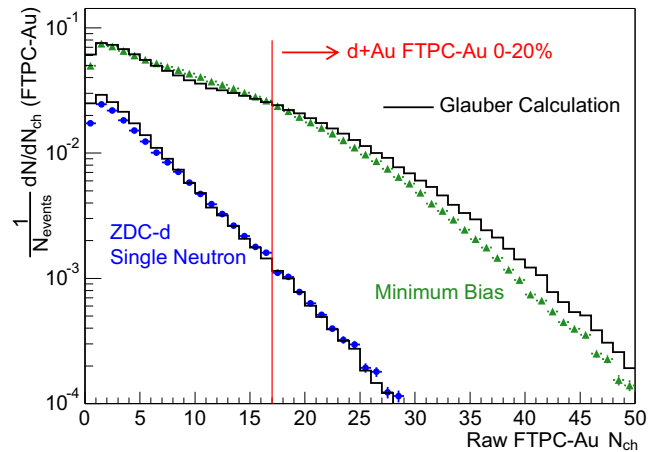}}
\caption{
Uncorrected charged particle multiplicity distributions measured in
$-3.8 < \eta < -2.8$ (Au-direction) for d+Au collisions. Points are
for minimum bias (triangles) and peripheral (circles, ZDC-d single neutron) collisions.
Both are normalized to the total number of d+Au collisions.
Histograms are Glauber model calculations.
\label{FigOne}}
\end{figure}

Centrality tagging of d+Au collisions is based on the raw
(uncorrected) charged particle multiplicity within $-3.8 < \eta <
-2.8$, measured by the Forward Time Projection Chamber in the Au
beam direction (FTPC-Au
\cite{STARNIM}). The FTPC-Au multiplicity was examined in quadrants 
relative to the orientation of the leading charged hadron at
mid-rapidity; auto-correlation effects were found to be
negligible. An independent centrality tag, used as a cross-check,
requires at least one beam-rapidity (spectator) neutron in ZDC-d, the
ZDC in the deuteron beam direction. The cross section for this process
in hadronic events was measured to be (19.2$\pm$1.3)\% of
\sigmadAuhadr.  ZDC-d and FTPC-Au are separated
by 8 rapidity units.
Figure \ref{FigOne} shows the FTPC-Au multiplicity for
minimum bias and ZDC-d neutron-tagged events.  The latter have a
strong bias toward low multiplicity.

The centrality tags were modeled using a Monte Carlo Glauber
calculation~\cite{STARHighpt130} incorporating the Hulth\'{e}n
wavefunction of the deuteron\cite{HulthenWF}. In this model the mean
number of binary collisions \NbinaryMean\ is 7.5$\pm0.4$ for minimum
bias events and
\sigmadAuhadr=2.21$\pm$0.09 b.
Events with a neutron spectator from the deuteron comprise
(18$\pm$3)\% of
\sigmadAuhadr\ in the model. This
event class is biased toward peripheral collisions, with
\NbinaryMean=2.9$\pm0.2$. The FTPC-Au multiplicity distribution was
modeled by convoluting the Glauber model distribution of participants from the Au
nucleus with the charged multiplicity
distribution measured in $2.5\lt|\eta|\lt3.5$ for
\pbar+p collisions at \sqrts=200 GeV\cite{UA5}. The
FTPC-Au acceptance, efficiency and backgrounds were taken into account
using HIJING
\cite{HIJING} events in a GEANT model of the detector. Figure
\ref{FigOne} shows the measurements for both minimum bias and ZDC-d
neutron-tagged events, together with the corresponding Glauber model
predictions. The model is validated by its agreement with both
multiplicity distributions and with the ZDC-d single neutron cross
section fraction. High FTPC-Au multiplicity therefore biases towards
central collisions. Figure
\ref{FigOne} shows the cut defining the 20\% highest multiplicity collisions in the data.
\NbinaryMean=15.0$\pm$1.1 for the 20\% highest multiplicity collisions in the 
Glauber model, where the uncertainty includes the spread in values obtained with several
alternative models.


\begin{figure}
\resizebox{\FigFactor\textwidth}{!}{\includegraphics{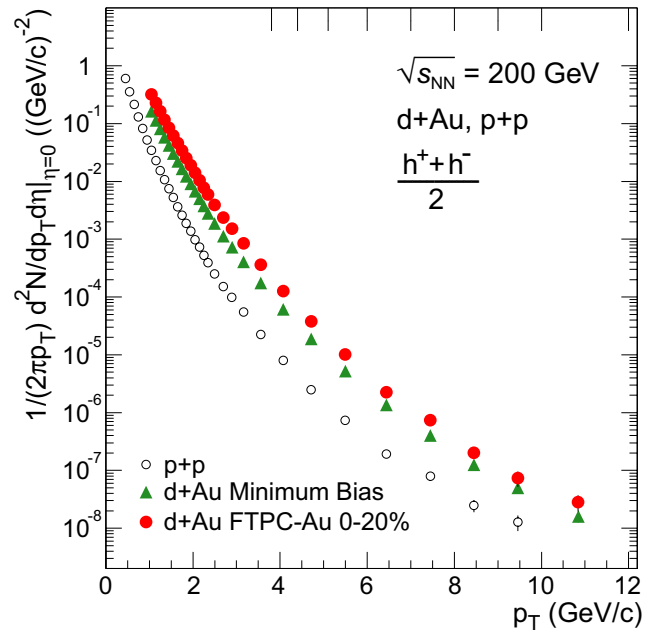}}
\caption{
Inclusive \pT\ distributions for minimum bias and central d+Au
collisions, and non-singly diffractive p+p collisions~\cite{STARHighpt200}. Hash marks at
the top indicate bin boundaries for \pT\gt3.8 GeV/c.
\label{FigTwo}}
\end{figure}

\begin{figure}
\resizebox{\FigFactor\textwidth}{!}{\includegraphics{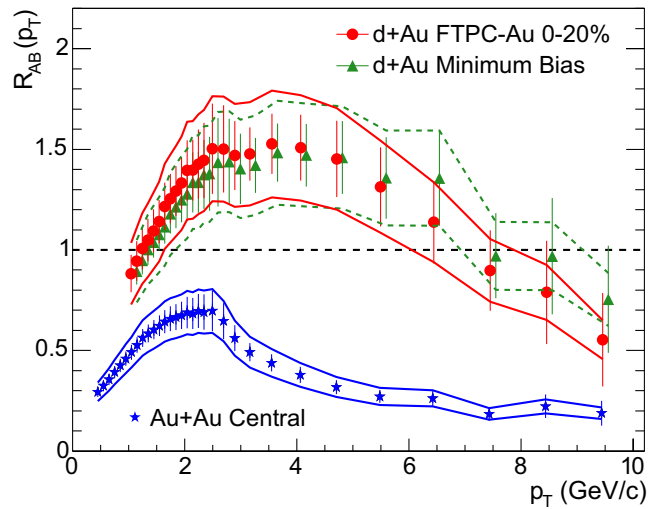}}
\caption{
\RAB\ from Eq.~\ref{RAA} for minimum bias 
and central d+Au collisions, and central Au+Au
collisions\cite{STARHighpt200}. The minimum bias d+Au data are
displaced 100 MeV/c to the right for clarity. The bands show the
normalization uncertainties, which are highly correlated
point-to-point and between the two d+Au distributions.
\label{FigThree}}
\end{figure}

Figure \ref{FigTwo} shows the invariant inclusive \pT\ distribution of
\hphm\ within $|\eta|\lt0.5$ for minimum bias and central d+Au
collisions, together with that for p+p collisions from
\cite{STARHighpt200}. The error bars are the quadrature sum of statistical errors and point-to-point 
systematic uncertainties. The normalization uncertainty for d+Au
collisions is 10\%.

Nuclear effects on hadron production in d+Au and Au+Au collisions are measured
through comparison to the p+p spectrum using the ratio
\begin{equation}
\label{RAA}
\RAB=\frac{d^2N/d{\pT}d\eta}{\TAB\,{d}^2\sigma^{pp}/d{\pT}d{\eta}}\ ,
\end{equation}
\noindent
where $d^2N/d{\pT}d\eta$ is the differential yield per event in the
nuclear collision $A+B$, \TAB=\NbinaryMean/\sigmappinel\ describes the 
nuclear geometry, and
${d}^2\sigma^{pp}/d{\pT}d{\eta}$ for p+p inelastic collisions is
determined from the measured p+p differential cross
section\cite{STARHighpt200}. In the absence of nuclear effects such as
shadowing, the Cronin effect, or gluon saturation, hard processes are expected to scale
with the number of binary collisions and \RAB=1.  Figure
\ref{FigThree} shows \RAB\ for minimum bias and central d+Au
collisions. The error bars are the quadrature sum of the statistical
and point-to-point systematic uncertainties. \RAB\gt1 for 2\lt\pT\lt7
GeV/c.
\RAB\ for central and minimum bias d+Au collisions
contain many common uncertainties, including dependence on the same p+p reference spectrum.
The ratio of \RAB\ for central relative to minimum bias collisions,
which factors out these common uncertainties, is 1.02$\pm$0.03 at 4 GeV/c.
\RAB\ may be influenced by nuclear shadowing~\cite{Eskola} 
and its centrality dependence~\cite{SpencerShadow}. Figure \ref{FigThree} also shows \RAB\ for
central Au+Au collisions\cite{STARHighpt200}, exhibiting large
suppression in hadron production at high \pT.


\begin{figure}
\resizebox{\FigFactor\textwidth}{!}{\includegraphics{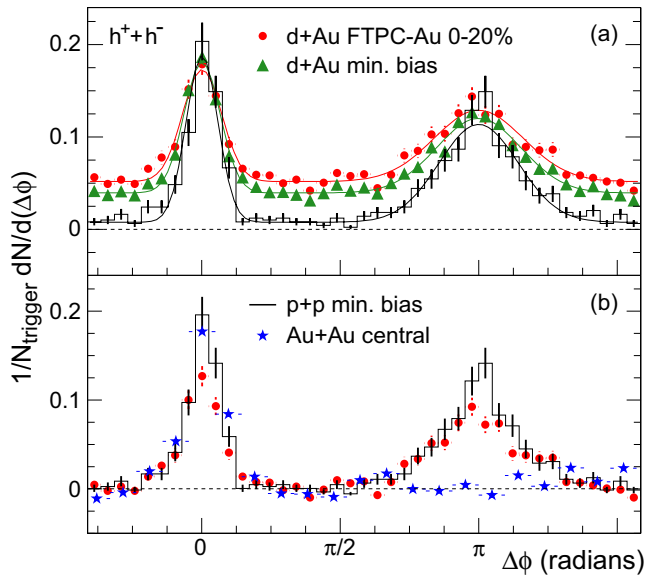}}
\caption{
(a) Efficiency corrected two-particle azimuthal distributions for minimum bias and central
d+Au collisions, and for p+p
collisions\cite{STARHighptCorrelations}. Curves are fits using
Eq.~\ref{EqCorr}, with parameters given in Table~\ref{TableCorr}.  (b)
Comparison of two-particle azimuthal distributions for central d+Au collisions to
those seen in p+p and central Au+Au collisions
\cite{STARHighptCorrelations}.  The respective pedestals have been
subtracted.
\label{FigFour}}
\end{figure}

Figure \ref{FigFour}(a) shows the two-particle azimuthal distribution
$D(\Delta \phi)$, defined as
\begin{equation} 
D(\Delta \phi) \equiv \frac{1}{N_{trigger}}\frac{1}{\epsilon}
\frac{dN}{d(\Delta \phi)} ,
\end{equation} 
for minimum bias and central d+Au collisions, and for p+p
collisions\cite{STARHighptCorrelations}. Only particles within
$|\eta|\lt0.7$ are included in the analysis. $N_{trigger}$ is the number
of particles within 4\lt\pT(trig)\lt6 GeV/c, referred to as trigger
particles. The distribution results from the correlation of each
trigger particle with all associated particles in the same event having
$2<p_T<p_T$(trig), where $\epsilon$ is the tracking efficiency of
the associated particles. The normalization uncertainties are less than 5\%.

\begin{table}
\caption{Fit parameters from Eq.~\ref{EqCorr}.  Errors are statistical only.
\label{TableCorr}}
\begin{ruledtabular}
\begin{tabular}{c|ccc}
                & p+p min.\@ bias         & d+Au min.\@ bias        & d+Au central          \\ \hline
$A_N$           & 0.081$\pm$0.005       & 0.073$\pm$0.003       & 0.067$\pm$0.004       \\
$\sigma_N$      & 0.18$\pm$0.01         & 0.20$\pm$0.01         & 0.22$\pm$0.02         \\
$A_B$           & 0.119$\pm$0.007       & 0.097$\pm$0.004       & 0.098$\pm$0.007       \\
$\sigma_B$      & 0.45$\pm$0.03         & 0.48$\pm$0.02         & 0.51$\pm$0.03         \\
$P$             & 0.008$\pm$0.001       & 0.039$\pm$0.001       & 0.052$\pm$0.002       \\
\end{tabular}
\end{ruledtabular}
\end{table}

The azimuthal distributions in d+Au collisions include a near-side
($\Delta\phi\sim0$) peak similar to that seen in p+p and
Au+Au collisions~\cite{STARHighptCorrelations} that is typical of jet
production, and a back-to-back ($\Delta\phi\sim\pi$) peak similar to
that seen in p+p and peripheral Au+Au
collisions~\cite{STARHighptCorrelations} that is typical of di-jet
events.
The azimuthal distributions are characterized by a fit to the sum of
near-side (first term) and back-to-back (second term) Gaussian peaks and a constant:
\begin{equation}
D(\Delta \phi)=
{A_N}\frac{e^{-(\Delta \phi)^2/2\sigma_N^2}}{\sqrt{2\pi}\sigma_N}+
{A_B}\frac{e^{-(|\Delta\phi|-\pi)^2/2\sigma_B^2}}{\sqrt{2\pi}\sigma_B}
+P .
\label{EqCorr}
\end{equation}
\noindent
Fit parameters are given in Table \ref{TableCorr}. Their systematic
uncertainties are highly correlated between the data sets, and are
less than 20\% for $\sigma_N$ and less than 10\% for all other
parameters. The only large difference in the
azimuthal distributions in p+p and d+Au collisions is the growth of
the pedestal $P$. It increases with increasing \NbinaryMean, but is
not proportional to
\NbinaryMean\ as might be expected for incoherent production. Both
$\sigma_N$ and $\sigma_B$ exhibit at most a small increase from p+p to
central d+Au collisions. A small growth in $\sigma_B$ is expected to
result from initial-state multiple scattering \cite{Corcoran,QiuVitev}.
The modest reduction in the correlation
strengths $A_N$ and $A_B$ from p+p to central d+Au
collisions is similar to that seen previously for
peripheral Au+Au collisions \cite{STARHighptCorrelations}.

Figure \ref{FigFour}(b) shows the pedestal-subtracted azimuthal
distributions for p+p and central d+Au collisions.
The azimuthal distributions are shown also for central Au+Au
collisions after subtraction of the elliptic flow and pedestal contributions
\cite{STARHighptCorrelations}. The near-side peak is
similar in all three systems, while the back-to-back peak in central
Au+Au shows a dramatic suppression relative to p+p and d+Au.


The contrast between d+Au and central Au+Au collisions in
Figs.~\ref{FigThree} and \ref{FigFour} indicates that the cause of the
strong high \pT\ suppression observed previously is associated with
the medium produced in Au+Au but not in d+Au collisions.  The
suppression of the inclusive hadron yield at high \pT\ in central
Au+Au collisions has been discussed theoretically in various
approaches (see \cite{STARHighpt200} for references). Measurements of
central Au+Au collisions \cite{STARHighpt200} are described both by
pQCD calculations that incorporate shadowing, the Cronin effect, and
partonic energy loss in dense matter, and by a calculation extending
the saturation model to high momentum transfer. However, predictions
of these models differ significantly for d+Au collisions.  Due to the
Cronin effect, pQCD models predict that \RAB\gt1 within $2\lt\pT\lt6$
GeV/c for minimum bias d+Au collisions, with a peak magnitude of 1.1-1.5
in the range $2.5\lt\pT\lt4$ GeV/c
\cite{Accardi}. The enhancement is expected to be larger for central
collisions \cite{VitevdAu}. The saturation model calculation in
\cite{KLM} predicts \RAB\lt1, with larger suppression
for more central events, achieving
\RAB$\sim0.75$ for the 20\% most central collisions. In contrast, another 
saturation model calculation~\cite{BaierEtalSaturation} generates an
enhancement in \RAB, similar to the Cronin effect, for both d+Au and Au+Au collisions.
Figure \ref{FigThree} shows that
\RAB\ is qualitatively different in d+Au and central Au+Au collisions: in
d+Au, \RAB\ significantly exceeds unity.  These results are consistent with
expectations from pQCD calculations but not the saturation model in
\cite{KLM}. Scattering of the hadronic fragments of jets 
also may contribute to the suppression of the inclusive
yield~\cite{GallmeisterEtAl,STARHighpt200}.

The azimuthal distributions of back-to-back jets and high \pT\
di-hadrons have been observed to broaden in fixed-target p+nucleus
collisions
relative to p+p collisions, but are not strongly suppressed
\cite{Corcoran}. Slight broadening of the back-to-back
hadron distribution in d+Au collisions at \sqrtsNN=200 GeV is also expected
from pQCD models incorporating the Cronin effect
\cite{QiuVitev}. Predictions of the
saturation model for the back-to-back hadron distributions require
further theoretical development, though the rate may be suppressed due
to a monojet contribution \cite{BtoBsaturation}.
Table \ref{TableCorr} shows that the distribution of back-to-back high
\pT\ hadrons is not substantially modified in central d+Au collisions
relative to p+p collisions, consistent with expectations from pQCD
calculations.

In summary, we have reported the inclusive \pT\ distributions and
two-particle azimuthal distributions of high \pT\ hadrons in minimum
bias and central d+Au collisions at \sqrtsNN=200 GeV. Similar
measurements for Au+Au and p+p interactions have revealed a striking
suppression of both the inclusive hadron yield and the back-to-back
correlations for central Au+Au collisions. If the suppression is the
result of initial-state effects, it also should be observed in d+Au
collisions. No suppression in d+Au collisions is observed. Rather, the
inclusive yield is enhanced and the two-particle azimuthal
distributions exhibit little change relative to p+p. These results
suggest that the Cronin effect plays a significant role in d+Au
collisions for $2<p_T<7$ GeV/c. We conclude that the
suppression phenomena seen in central Au+Au collisions are due to
final-state interactions with the dense system generated in the
collision.


\begin{acknowledgments}
We thank the RHIC Operations Group and RCF at BNL, and the
NERSC Center at LBNL for their support. This work was supported
in part by the HENP Divisions of the Office of Science of the U.S.
DOE; the U.S. NSF; the BMBF of Germany; IN2P3, RA, RPL, and
EMN of France; EPSRC of the United Kingdom; FAPESP of Brazil;
the Russian Ministry of Science and Technology; the Ministry of
Education and the NNSFC of China; SFOM of the Czech Republic,
DAE, DST, and CSIR of the Government of India; the Swiss NSF.
\end{acknowledgments}


\def\etal{\mbox{$\mathrm{\it et\ al.}$}}

\end{document}